\documentclass[12pt,final]{amsart}

\usepackage{graphicx} 
\usepackage[margin=0.7in]{geometry}
\usepackage{amsmath}
\usepackage{nopageno}

\usepackage{verbatim}
\usepackage{amssymb}
\usepackage{amscd}
\usepackage{circuitikz}
\usepackage{ulem}

\newtheorem{lemma}{Lemma}[section]
\newtheorem{proposition}[lemma]{Proposition}
\newtheorem{theorem}[lemma]{Theorem}

\theoremstyle{definition}

\theoremstyle{remark}
\newtheorem{remark}[lemma]{Remark}
\newtheorem{example}[lemma]{Example}

\newcommand{\N}{\ensuremath{{\mathbb N}}}

\newcommand{\R}{\ensuremath{\mathbb R}}

\author{Tatyana Barron}
\address{Department of Mathematics, University of Western Ontario, London Ontario N6A 5B7, Canada}
\email{tatyana.barron@uwo.ca}

\title{Mathematical models for therapeutic approaches involving electric conductors or shielding}
\begin{document}
\sloppy

\maketitle

\noindent {\bf Abstract}. We set up a 
mathematical model for a DC current in a human tissue that shows an attenuation effect in an extended circuit. We give a positive lower bound on the time duration over which this is guaranteed to happen in terms of the parameters of the model. We also discuss shielding and coupling  
in the context of electrical aspects of biological processes. 

\

\noindent {\bf Keywords}: ordinary differential equations, signal propagation, neurons,  biomedical circuits 

\

\noindent {\bf 2020 MSC}: 92-10, 92C50

\section{Introduction}
\label{sec:intro}

Transmission of electrical signals plays an important part in biological processes. It is a basis for many diagnostic, therapeutic, and biomedical applications, including, to give some specific examples, ECG \cite{gesel} or  pulsed radio frequency energy therapy  \cite{mtassone}, \cite{mirpuri}, \cite{sorrell},  \cite{tassone}.  

The mathematical methods of neuroscience often address the propagation of signals in networks of neurons in humans and related aspects of data analysis. From a somewhat different perspective, one can  consider, abstractly, 
parallels between a human body and computer hardware. This is the point of view echoed in the ideas around the First draft \cite{vonn}, \cite{biocomput}. Certainly, humans are more complex than computational systems \cite{penrose}.  There are chemical aspects of signal propagation (e.g. binding of neurotransmitters to receptors). The mechanism of propagation of electical signals through neurons is not the same as that of an electrical current in a conductor wire, even though it is modeled, mathematically, by an electrical circuit \cite{hodghux}. 
One can  view propagation of electrical signals  
in nerves as pulsed DC current. 
More generally, 
there are endogenic DC currents and electric fields everywhere in the body \cite{gesel}, \cite{mh}. Nerves, specifically, can be modeled as electrical cable (\cite{mh} section 5.4). 
It is customary to represent these processes by electrical circuits: such as Fig. 1 in \cite{hodghux} or 
Fig. 5.15 in \cite{mh}, see also Fig. 9.6 and discussion in section 9.2.4.1 \cite{mh}, or, as another example, 
 Fig. 8 in \cite{skibret}. The endogenic electric currents induce magnetic fields  (section 7.12.8.3 of \cite{mh}, also \cite{brenner}, 
\cite{cohen}, \cite{yama}). In turn, magnetic fields affect biological processes. Experimental evidence, mechanisms, and 
therapeutic uses are discussed, in particular,  in     \cite{statm}, \cite{funk}, 
     \cite{aielloplus}, \cite{prato},  \cite{rthesis},  \cite{pratoetc},  \cite{valberg},  \cite{kirsch},  \cite{simon23},  \cite{simon22}.    
    Electromagnetic  induction plays a role. This is aligned with the fact well known in chemistry, that the magnetic fields can influence rates and yields of chemical reactions \cite{rodgers}, 
\cite{stass},  and, from a broader perspective, with the general understanding that the processes in human body, including those in the nervous system, are not a closed system and the models should include the environment (see e.g. the discussion 
in Chapter 14 \cite{sporns}).  

In this paper, we consider the question of pain reduction in humans through the lens of propagation of electric signals in nerves. This is only a model or an approximation. We leave any medical conclusions to the clinical experts. 
Our 
Theorems \ref{thelectrod}, \ref{thelectrodgen} show that extending the circuit (Section  \ref{secextc}) reduces the electric current in the nerve. Part (b) in each of these theorems gives a time interval $[0,T)$ with a positive lower bound on $T$, over which this is guaranteed to happen. This $T$ is not the best possible bound. In Sections \ref{interf}, \ref{sec:wifi}  we discuss, in this context,  shielding and coupling which are, in the standard signal transmission,  "passive cancel" techniques for reducing interference.

\section{Extending the circuit} 
\label{secextc}
\begin{figure}[tb]
\centering
\includegraphics[width=2.5in]{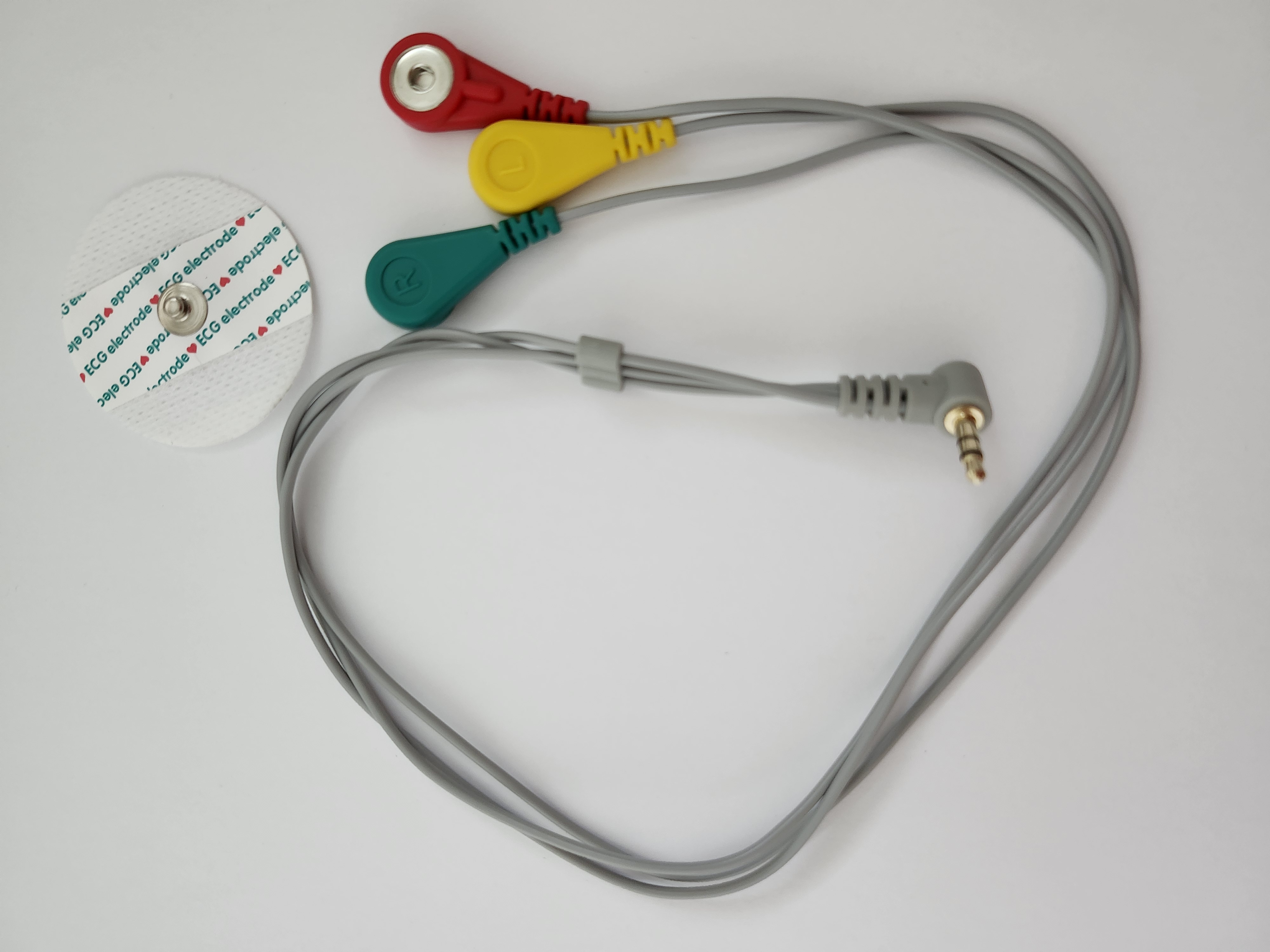}
\caption{A standard Ag/AgCl electrode and an electrode wire.} \label{figelec}
\end{figure}

We suggest that placing two or more pick up Ag/AgCl pick up electrodes (Fig. \ref{figelec}) connected with each other by an electrode wire, at the pain site (e.g. in case of a minor headache that seems to originate close to the surface of the skin)   
would have a therapeutic effect. There is a biopotential difference between two contact sites (\cite{mh} Chapter 7). 
See also the discussion in \cite{gedval}. 

 Below, we set up a mathematical model and conclude (Theorems \ref{thelectrod}, \ref{thelectrodgen}) that placing the electrodes attenuates  the electrical signal. 
 
\subsection{Mathematical model} 
\label{sec:mmodel}
Instead of the circuit in Fig. 1 of \cite{hodghux}, to model the set up, we will consider 
the simplest RC circuit, pictured in Figure \ref{circt1}, with a direct current voltage source $V$, a resistor $R$ and a capacitor $C$.  
We represent adding two reference electrodes with a wire between them by adding an $R| \ |C$ circuit: Figure \ref{circt2}.

\begin{figure}
\begin{minipage}{0.35\textwidth}
\centering
\begin{circuitikz}
\draw   (0,0)   coordinate(start)              
                to [short, i=$I$] ++ (1,0)    
                to [R=$R$ ]      ++ (3,0)    coordinate (a) 
                -- ++ (0,-1)
                to [C=$C$]    ++ (-3,0)
                -- ++ (-1,0)
                to [battery, l=$V$, invert] (start)
        ;
\end{circuitikz}
\caption{Circuit 1: human tissue, no electrodes} \label{circt1}
\end{minipage}
\begin{minipage}{0.35\textwidth}
\centering
\begin{circuitikz}
\draw   (0,0)   coordinate(start) 
                to [short, i=$I$] ++ (1,0)    
                to [R=$R$ ]      ++ (3,0)    coordinate (a) 
                to [short, -*]      ++ (0,-1)   coordinate (b)
                to [short, i_=$I_1$]    ++ (-1.2,0)
                to [R=$R_e$]      ++ (0,-2)
                to [short, -*]      ++ (1.2,0)  coordinate (c)
        (b)     to [short, i=$I_2$]         ++ (+1.2,0)
                to [C=$C_e$]    ++ (0,-2) -- (c)
                -- ++ (0,-1)
                to [C=$C$]    ++ (-3,0)
                -- ++ (-1,0)
                to [battery, l=$V$, invert] (start)
        ;
\end{circuitikz}
\caption{Circuit 2: human tissue and two attached reference electrodes connected with a wire} \label{circt2}
\end{minipage}
\end{figure}

We apply the Kirchhoff laws. 
For circuit $1$, it is well known that 
 for small values of time $t>0$ 
 the capacitor charge
 $$ 
 q(t)=VC(1-e^{-\frac{1}{RC}t})
 $$
 and the current is 
 \begin{equation}
 \label{itc1}
 \frac{dq}{dt}=I(t)=\frac{V}{R}e^{-\frac{1}{RC}t}.
 \end{equation}
 For circuit $2$, we obtain a system of equations: 
 $$
\left\{ 
\begin{array}{lllll }  
V-IR-I_1R_e-\frac{q}{C}=0 \\
V-IR-\frac{q_e}{C_e}-\frac{q}{C}=0 \\
I_1+I_2=\dfrac{dq}{dt}  \\
I_2=\dfrac{dq_e}{dt} \\
I=I_1+I_2 
\end{array} \right. 
$$
with initial conditions $q(0)=0$ and $q_e(0)=0$. 
Here $C_e$ and $R_e$ are the  capacitance and resistance of the electrodes/multi-wire system, respectively, 
 and $q_e$ is the charge on the second capacitor (added in circuit 2). 
Note: $I_1R_e=\frac{q_e}{C_e}$. 
From the equations, we get  the second order ODE with constant coefficients 
\begin{equation}
\label{eqode}
q_e''+q_e'(\frac{1}{R_eC_e}+\frac{1}{RC}+\frac{1}{RC_e})+q_e\frac{1}{RCR_eC_e}=0.
\end{equation}
This equation, with the initial conditions above, has the solution 
\begin{equation}
\label{qt}
q_e(t)=\frac{V}{R(\lambda_1-\lambda_2)}(e^{\lambda_1t}-e^{\lambda_2t})
\end{equation}
 where the two distinct negative solutions of the equation 
 $$
\lambda^2+\lambda (\frac{1}{R_eC_e}+\frac{1}{RC}+\frac{1}{RC_e})+\frac{1}{RCR_eC_e}=0
 $$
 are
 $$
 \lambda_1=\frac{1}{2}\Bigl ( -(\frac{1}{R_eC_e}+\frac{1}{RC}+\frac{1}{RC_e})+
 \sqrt{(\frac{1}{R_eC_e}+\frac{1}{RC}+\frac{1}{RC_e})^2-\frac{4}{RCR_eC_e} } \Bigr ) ,
 $$
 $$
 \lambda_2=\frac{1}{2}\Bigl ( -(\frac{1}{R_eC_e}+\frac{1}{RC}+\frac{1}{RC_e})-
 \sqrt{(\frac{1}{R_eC_e}+\frac{1}{RC}+\frac{1}{RC_e})^2-\frac{4}{RCR_eC_e} } \Bigr ) .
 $$
 From (\ref{qt}) 
 \begin{equation}
 \label{itc2}
 I(t)=I_1(t)+I_2(t)=\frac{V}{R(\lambda_1-\lambda_2)}
 \Bigl ( 
 \frac{1}{R_eC_e}(e^{\lambda_1t}-e^{\lambda_2t})+(\lambda_1e^{\lambda_1t}-\lambda_2e^{\lambda_2t}) \Bigr ).
 \end{equation}
 Observe: 
 $$
 \lambda_2<\lambda_1<0; \ \lambda_1-\lambda_2>0; \ 
  -\frac{1}{\lambda_1}> -\frac{1}{\lambda_2}>0. 
  $$
 At $t=0$, the currents (\ref{itc1}) and  (\ref{itc2}) have the same value $\frac{V}{R}$. 
 \begin{theorem} Assume all notations are as above.  
 \label{thelectrod}
 
 (a) There is $T>0$ such that for all $t\in (0,T)$ the current (\ref{itc1}) is strictly larger than the current (\ref{itc2}). 
 
 (b) The statement $(a)$ holds, in particular, for $T=\tau$, where 
 $$
 \tau=\min \{ 
 \frac{2}{\frac{C_e}{R}(\frac{1}{C}+\frac{1}{C_e})^2+\frac{1}{R_eC_e}};RC; 
 -\frac{1}{\lambda_2}\} >0.
 $$
\end{theorem}
 {\bf Proof.} 
 The function 
 $$
 f(t)= \frac{V}{R}e^{-\frac{1}{RC}t}-
 \frac{V}{R(\lambda_1-\lambda_2)}\Bigl ( 
 \frac{1}{R_eC_e}(e^{\lambda_1t}-e^{\lambda_2t})+(\lambda_1e^{\lambda_1t}-\lambda_2e^{\lambda_2t}) \Bigr )
; \ t\ge 0
 $$
 is $C^{\infty}$, \ $f(0)=0$, and 
 $$
 f'(0)=
  \frac{V}{R^2C_e} 
 >0. 
 $$
 By continuity of $f'(t)$ it follows that $f'(t)>0$ in the interval $(0,T)$ for some $T$, meaning that $f$ is increasing in this interval, and since $f(0)=0$, it follows that $f(t)>0$ in $(0,T)$. This proves $(a)$. 
 
Now we will prove $(b)$. Consider the Taylor series at $t=0$ for $y=\frac{V}{R}e^{-\frac{1}{RC}t}$. It converges for all $t$. 
For each $t$ in the interval $(0,RC)$, it is an alternating series of the form $\sum\limits_{k=0}^{\infty} (-1)^kb_k$, where $b_k>0$ for each $k$ 
and the sequence $(b_k)$ is decreasing.  
Denote by $P_1(t)$ and $P_2(t)$ the degree $1$ and $2$ Taylor polynomials of this series: 
$$
P_1(t)=\frac{V}{R}(1-\frac{1}{RC}t); \ 
P_2(t)=\frac{V}{R}(1-\frac{1}{RC}t+\frac{1}{2}\frac{1}{(RC)^2}t^2).
$$
Consider the Taylor series at $t=0$ for 
\begin{equation}
\label{secondf}
y=
 \frac{V}{R(\lambda_1-\lambda_2)}\Bigl ( 
 \frac{1}{R_eC_e}(e^{\lambda_1t}-e^{\lambda_2t})+(\lambda_1e^{\lambda_1t}-\lambda_2e^{\lambda_2t}) \Bigr ). 
\end{equation}  
This Taylor series, for each $t$ in the interval $(0, -\frac{1}{\lambda_2})$, 
is an alternating series of the form $\sum\limits_{k=0}^{\infty} (-1)^kb_k$, where $b_k>0$ for each $k$ 
and the sequence $(b_k)$ is decreasing.  
 Denote by $Q_1(t)$ and $Q_2(t)$ the degree $1$ and $2$ Taylor polynomials of this series: 
$$
Q_1(t)=\frac{V}{R}[\frac{1}{R_eC_e}t+1+(\lambda_1+\lambda_2)t]
$$
$$
Q_2(t)=\frac{V}{R}[\frac{1}{R_eC_e}t+\frac{1}{2R_eC_e}(\lambda_1+\lambda_2)t^2+1+(\lambda_1+\lambda_2)t+ 
\frac{1}{2}(\lambda_1^2+\lambda_1\lambda_2+\lambda_2^2)t^2].
$$
We conclude: for each $t$ in the interval $[0, 
\min \{-\frac{1}{\lambda_2};RC\} )$
$$
P_1(t)\le \frac{V}{R}e^{-\frac{1}{RC}t}\le P_2(t)
$$
and 
$$
Q_1(t)\le 
 \frac{V}{R(\lambda_1-\lambda_2)}\Bigl ( 
 \frac{1}{R_eC_e}(e^{\lambda_1t}-e^{\lambda_2t})+(\lambda_1e^{\lambda_1t}-\lambda_2e^{\lambda_2t}) \Bigr )
\le Q_2(t).
 $$
 Hence 
 $$
 P_1(t)-Q_2(t)\le f(t)\le P_2(t)-Q_1(t), \ 0\le t <
\min \{-\frac{1}{\lambda_2};RC\}  .
 $$
 The quadratic polynomial 
 $$
P_1(t)-Q_2(t)=\frac{V}{R}(1-\frac{1}{RC}t)-
\frac{V}{R}[\frac{1}{R_eC_e}t+\frac{1}{2R_eC_e}(\lambda_1+\lambda_2)t^2+1+(\lambda_1+\lambda_2)t+ 
\frac{1}{2}(\lambda_1^2+\lambda_1\lambda_2+\lambda_2^2)t^2]
$$
has two nonnegative distinct roots: 
$$
t=0; \ t= 
\frac{2}{C_e}\frac{1}{\frac{1}{R}(\frac{1}{C}+\frac{1}{C_e})^2+\frac{1}{R_eC_e^2}}
$$
and takes positive values on the interval $(0,
\frac{2}{C_e}\frac{1}{\frac{1}{R}(\frac{1}{C}+\frac{1}{C_e})^2+\frac{1}{R_eC_e^2}}
)$ (because  
$P'_1(0)-Q'_2(0)=
\frac{V}{R^2C_e} >0$). Therefore $f(t)>0$ on the interval $(0,\tau)$, and the statement $(b)$ follows. $\Box$

\begin{figure}[tb]
\includegraphics[width=3.2in]{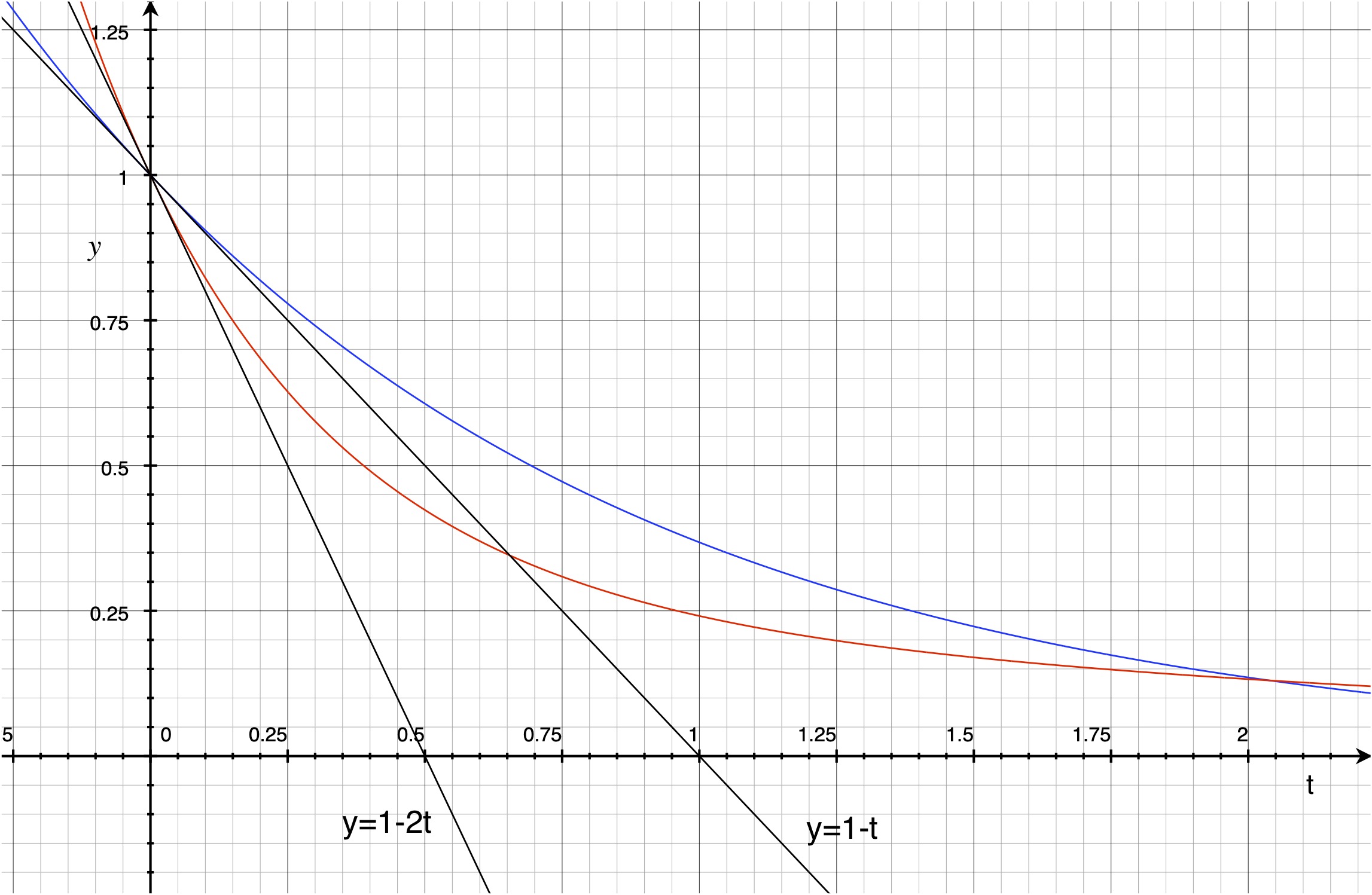}
\caption{
The currents and linear approximations for Example \ref{ex2circt}. The blue curve is $y=e^{-t}$. 
The red curve is the graph of (\ref{theotherc}). The lines are the graphs of  $P_1(t)=1-t$ and $Q_1(t)=1-2t$. }
\label{figgra}
\end{figure}

\begin{figure}[tb]
\includegraphics[width=3.2in]{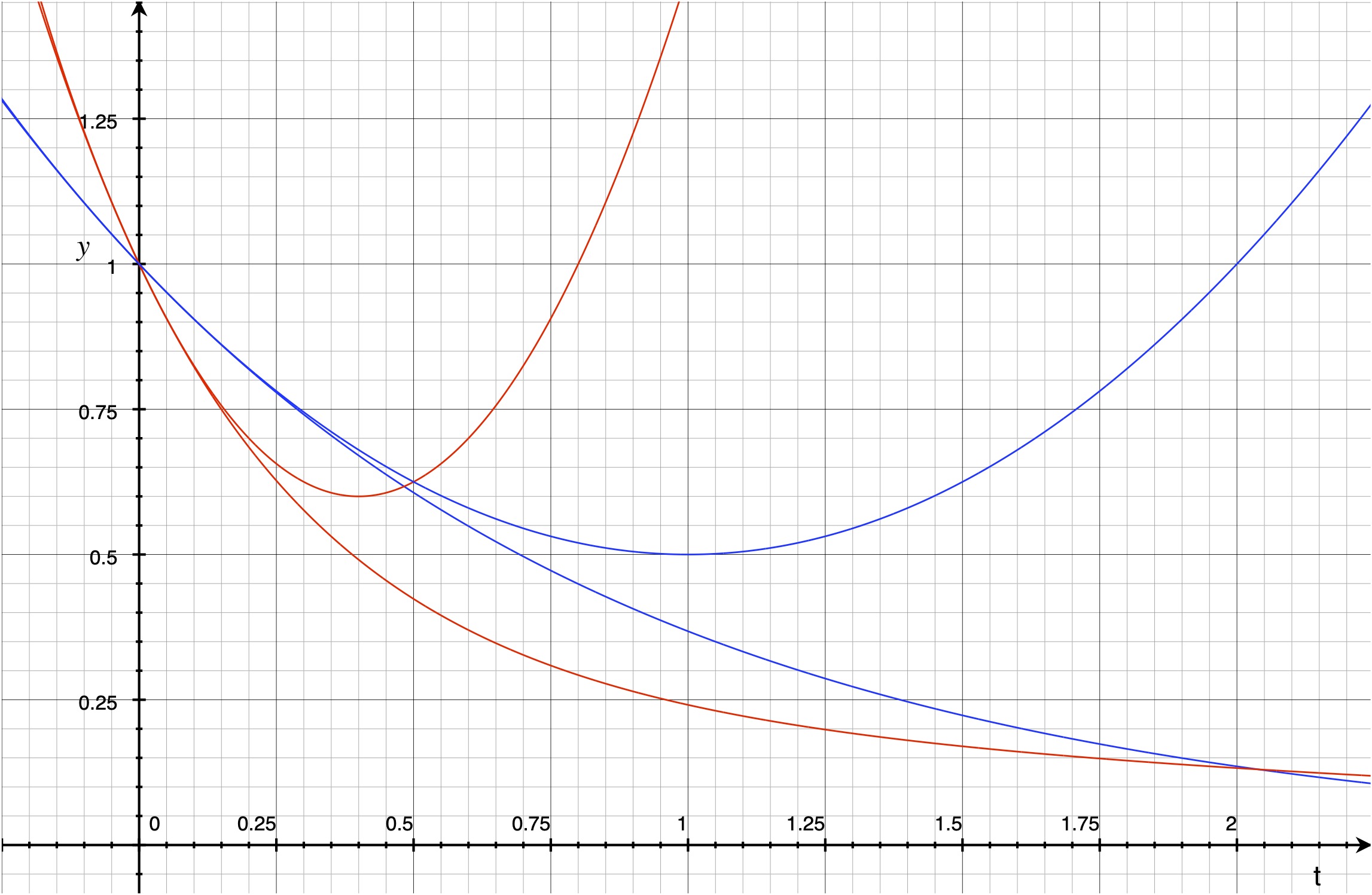}
\caption{The currents and quadratic approximations for Example \ref{ex2circt}. The blue curves are the graphs of  
$y=e^{-t}$ and its degree $2$ Taylor polynomial $P_2$. 
The red curves are the graphs of (\ref{theotherc}) and its degree $2$ Taylor polynomial 
$Q_2$.  } \label{figquadr}
\end{figure}

\begin{figure}[tb]
\includegraphics[width=3.2in]{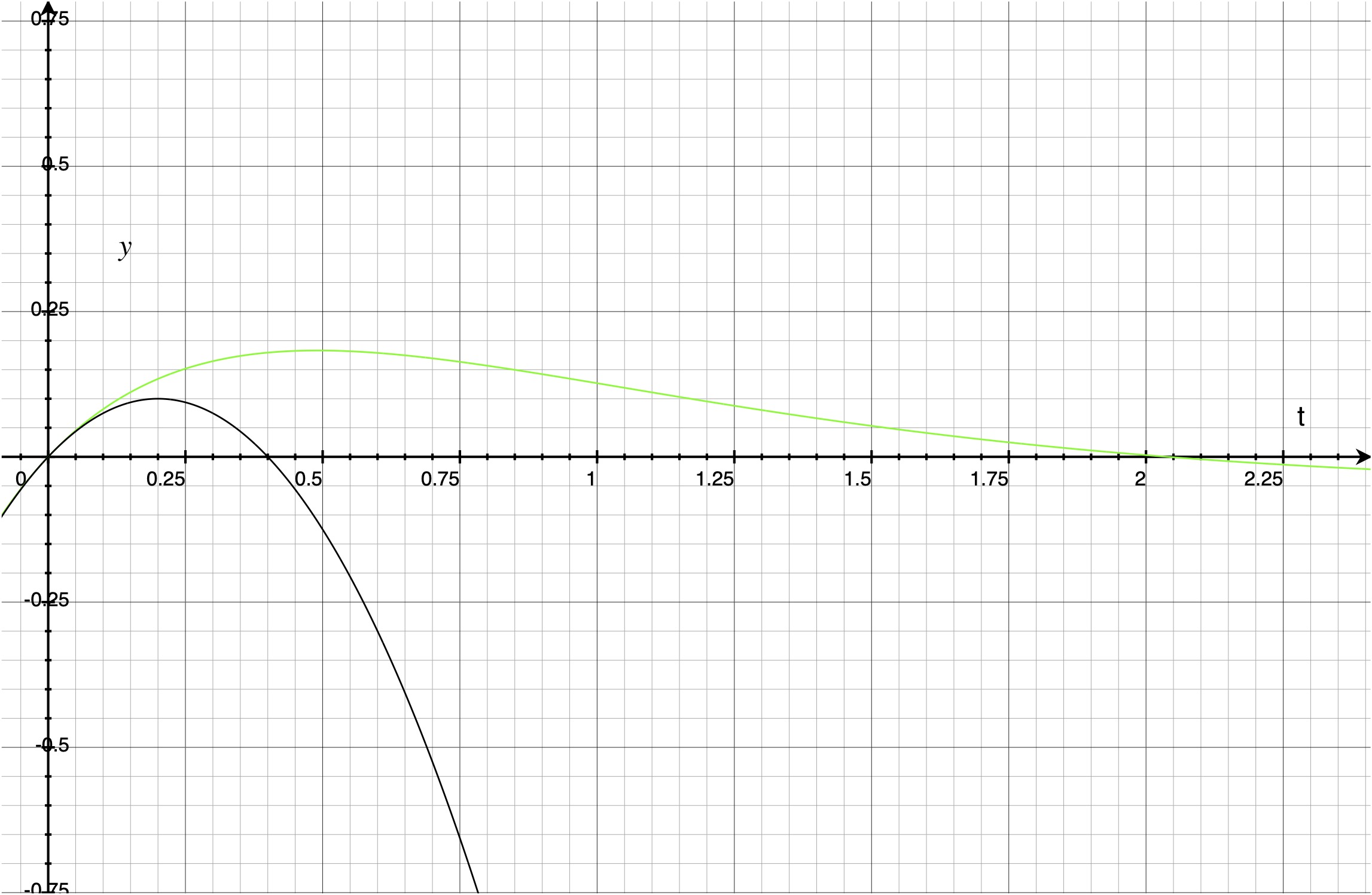}
\caption{The green curve is the graph of  $y=f(t)$. 
The black curve is the graph of $y=P_1(t)-Q_2(t)=(1-t)-(1-2t+2.5t^2)$. } \label{figpq}
\end{figure}

The positive bound in the part $(b)$ of Theorem \ref{thelectrod} is not optimal. We illustrate the Theorem with the example below. 
 \begin{example}
 \label{ex2circt}
 Let us compare the currents for circuit 1 and circuit 2 (equations (\ref{itc1}) and (\ref{itc2})) for 
 $V=1$, $C=1$, $R=1$, 
 $C_e=1$, $R_e=1$ (we are omitting the units). 
 The current for circuit $1$, given by equation (\ref{itc1}) is $I(t)=e^{-t}$. 
 The current for circuit $2$, given by equation (\ref{itc2}) (noting that $\lambda_{1}=-1.5+0.5\sqrt{5}$, \ 
 $\lambda_{2}=-1.5-0.5 \sqrt{5}$) is 
 \begin{equation}
 \label{theotherc}
 I(t)=\frac{1}{\sqrt{5}}\Bigl ( (-0.5+0.5\sqrt{5})e^{(-1.5+0.5\sqrt{5})t} +(0.5+0.5\sqrt{5})e^{(-1.5-0.5\sqrt{5})t} \Bigr ). 
\end{equation}
 The linear approximations for these two functions at $t=0$ are $P_1(t)=1-t$ and $Q_1(t)=1-2t$, respectively. 
 The quadratic approximations are $P_2(t)=1-t+0.5t^2$ and $Q_2(t)=1-2t+2.5t^2$. 
 Referring to Theorem \ref{thelectrod} and its proof, 
 $$
 f(t)=e^{-t}-\frac{1}{\sqrt{5}}\Bigl ( (-0.5+0.5\sqrt{5})e^{(-1.5+0.5\sqrt{5})t} +(0.5+0.5\sqrt{5})e^{(-1.5-0.5\sqrt{5})t} \Bigr )
 $$
 the polynomial 
 $$
 P_1(t)-Q_2(t)=t-2.5t^2
 $$
 has roots $0$ and $0.4$. See Fig. \ref{figgra}, \ref{figquadr}, \ref{figpq}.  
 The interval provided in Theorem \ref{thelectrod} $(b)$ is $(0,\tau)$ where 
 $$
 \tau=\min\{ 0.4; 1;\frac{2}{3+\sqrt{5}}\}=\frac{2}{3+\sqrt{5}}=\frac{2(3-\sqrt{5})}{9-5}=1.5-0.5\sqrt{5}. 
 $$
 \end{example}

\subsection{Further analysis} 
\label{sec:furan}

Let us extend the mathematical model in section \ref{sec:mmodel} to $k+1$ electrodes, where $k$ is a positive integer, and examine our argument to see how generalizing from $k=1$ to an arbitrary $k\ge 1$ affects the conclusion.  

We will model the electrodes and wires by $k$ $R||C$ blocks connected in series, and we will slightly change notation for the electrical currents. We generalize the circuit in   
Fig. \ref{circt2} to the circuit in Fig. \ref{kelectr}. 

\begin{figure}[tb]
\centering
\begin{circuitikz}
\draw   (0,0)   coordinate(start) 
                to [short, i=$I$] ++ (1,0)    
                to [R=$R$ ]      ++ (3,0)    coordinate (a) 
                to [short, -*]      ++ (0,-1)   coordinate (b)
                to [short, i_=$I_1$]    ++ (-1.2,0)
                to [R=$R_e$]      ++ (0,-2)
                to [short, -*]      ++ (1.2,0)  coordinate (c)
        (b)     to [short, i=$\tilde{I}_1$]         ++ (+1.2,0)
                to [C=$C_e$]    ++ (0,-2) -- (c)
                -- ++ (0,-1)
       
                       to [short, -*]      ++ (0,-1)   coordinate (b)         
                        to [short, i_=$I_2$]    ++ (-1.2,0)
                to [R=$R_e$]      ++ (0,-2)
                to [short, -*]      ++ (1.2,0)  coordinate (c)
        (b)     to [short, i=$\tilde{I}_2$]         ++ (+1.2,0)
                to [C=$C_e$]    ++ (0,-2) -- (c)
                -- ++ (0,-1)
                
                         to [short, -*]      ++ (0,-1)   coordinate (b)         
                        to [short, i_=$I_3$]    ++ (-1.2,0)
                to [R=$R_e$]      ++ (0,-2)
                to [short, -*]      ++ (1.2,0)  coordinate (c)
        (b)     to [short, i=$\tilde{I}_3$]         ++ (+1.2,0)
                to [C=$C_e$]    ++ (0,-2) -- (c)
                -- ++ (0,-1)

                to [C=$C$]    ++ (-3,0)
                -- ++ (-1,0)
                to [battery, l=$V$, invert] (start)
        ;
\end{circuitikz}
\caption{General case: $k+1$ electrodes (in this diagram, $k=3$).} \label{kelectr}
\end{figure}
We obtain the following equations. 
$$
\left\{ 
\begin{array}{llllll }  
V-IR-(I_1+...+I_k)R_e-\frac{q}{C}=0 \\
I_iR_e=\frac{q_i}{C_e}; \ i\in \{1,...,k\} \\
I_i+\tilde{I}_i=I=\dfrac{dq}{dt}; \  i\in \{1,...,k\}\\
\tilde{I}_i=\dfrac{dq_i}{dt}; \ i\in \{1,...,k\} \\
q(0)=0 \\ 
q_i(0)=0; \ i\in \{1,...,k\}
\end{array} \right. .
$$
Here $V$, $I$, $R$, $C_e$, $R_e$, $q$ are as above in section \ref{sec:mmodel}. The currents $I_i$, $\tilde{I}_i$ are as shown in Fig. \ref{kelectr}, and 
the charge $q_i$ is on the capacitor in the $i$-th component. 

From the equations, we get: 
$$
\frac{1}{R_eC_e}q_1+\frac{dq_1}{dt}=\frac{1}{R_eC_e}q_2+\frac{dq_2}{dt}; \ q_1(0)=q_2(0)=0.
$$
Solving for $q_2-q_1$, we conclude: $q_1(t)=q_2(t)$, $t\ge 0$. Similarly $q_i(t)=q_1(t)$ for every $i$. The first equation becomes 
$$
V-IR-\frac{k}{C_e}q_1-\frac{q}{C}=0.
$$
Differentiating this equation, we arrive at the following ODE (compare to (\ref{eqode}) in Sec. \ref{sec:mmodel}):
$$
q_1''+q_1'(\frac{1}{RC}+\frac{1}{R_eC_e}+\frac{k}{RC_e})+q_1\frac{1}{RCR_eC_e}=0.
$$
This equation, with the initial conditions above, has the solution 
\begin{equation}
\label{qtgen}
q_1(t)=\frac{V}{R(\lambda_1-\lambda_2)}(e^{\lambda_1t}-e^{\lambda_2t})
\end{equation}
 where the two distinct negative solutions of the equation 
 $$
\lambda^2+\lambda (\frac{1}{R_eC_e}+\frac{1}{RC}+\frac{k}{RC_e})+\frac{1}{RCR_eC_e}=0
 $$
 are
 $$
 \lambda_1=\frac{1}{2}\Bigl ( -(\frac{1}{R_eC_e}+\frac{1}{RC}+\frac{k}{RC_e})+
 \sqrt{(\frac{1}{R_eC_e}+\frac{1}{RC}+\frac{k}{RC_e})^2-\frac{4}{RCR_eC_e} } \Bigr ) ,
 $$
 $$
 \lambda_2=\frac{1}{2}\Bigl ( -(\frac{1}{R_eC_e}+\frac{1}{RC}+\frac{k}{RC_e})-
 \sqrt{(\frac{1}{R_eC_e}+\frac{1}{RC}+\frac{k}{RC_e})^2-\frac{4}{RCR_eC_e} } \Bigr ) .
 $$
 From (\ref{qtgen}) 
 \begin{equation}
 \label{itc2gen}
 I(t)=\frac{1}{R_eC_e}q_1(t)+\frac{dq_1}{dt}=\frac{V}{R(\lambda_1-\lambda_2)}
 \Bigl ( 
 \frac{1}{R_eC_e}(e^{\lambda_1t}-e^{\lambda_2t})+(\lambda_1e^{\lambda_1t}-\lambda_2e^{\lambda_2t}) \Bigr ).
 \end{equation}
 Observe: 
 $$
 \lambda_2<\lambda_1<0; \ \lambda_1-\lambda_2>0; \ 
  -\frac{1}{\lambda_1}> -\frac{1}{\lambda_2}>0. 
  $$
  At $t=0$, the currents (\ref{itc1}) and  (\ref{itc2gen}) have the same value $\frac{V}{R}$. 
 \begin{theorem} Assume all notations are as above.  
 \label{thelectrodgen}
 
 (a) There is $T>0$ such that for all $t\in (0,T)$ the current (\ref{itc1}) is strictly larger than the current (\ref{itc2gen}). 
 
 (b) The statement $(a)$ holds, in particular, for $T=\tau$, where 
 $$
 \tau=\min \{ 
 \frac{2k}{\frac{C_e}{R}(\frac{1}{C}+\frac{k}{C_e})^2+\frac{k}{R_eC_e}};RC; 
 -\frac{1}{\lambda_2}\} >0.
 $$
\end{theorem}
 {\bf Proof.} 
 The function 
 $$
 f(t)= \frac{V}{R}e^{-\frac{1}{RC}t}-
 \frac{V}{R(\lambda_1-\lambda_2)}\Bigl ( 
 \frac{1}{R_eC_e}(e^{\lambda_1t}-e^{\lambda_2t})+(\lambda_1e^{\lambda_1t}-\lambda_2e^{\lambda_2t}) \Bigr )
; \ t\ge 0
 $$
 is $C^{\infty}$, \ $f(0)=0$, and 
 $$
 f'(0)=
 k \frac{V}{R^2C_e} 
 >0. 
 $$
 By continuity of $f'(t)$ it follows that $f'(t)>0$ in the interval $(0,T)$ for some $T$, meaning that $f$ is increasing in this interval, and since $f(0)=0$, it follows that $f(t)>0$ in $(0,T)$. This proves $(a)$. 
 
The proof of  $(b)$ is similar to the proof of Theorem \ref{thelectrod} $(b)$:  
$$
P_1(t)\le \frac{V}{R}e^{-\frac{1}{RC}t}\le P_2(t)
$$
for all $t$ in the interval $(0,RC)$, 
where
$$
P_1(t)=\frac{V}{R}(1-\frac{1}{RC}t); \ 
P_2(t)=\frac{V}{R}(1-\frac{1}{RC}t+\frac{1}{2}\frac{1}{(RC)^2}t^2).
$$
Furthermore, 
$$
Q_1(t)\le 
 \frac{V}{R(\lambda_1-\lambda_2)}\Bigl ( 
 \frac{1}{R_eC_e}(e^{\lambda_1t}-e^{\lambda_2t})+(\lambda_1e^{\lambda_1t}-\lambda_2e^{\lambda_2t}) \Bigr )
\le Q_2(t)
 $$
 for each $t$ in the interval $(0, -\frac{1}{\lambda_2})$, where 
 $$
Q_1(t)=\frac{V}{R}[\frac{1}{R_eC_e}t+1+(\lambda_1+\lambda_2)t]
$$
$$
Q_2(t)=\frac{V}{R}[\frac{1}{R_eC_e}t+\frac{1}{2R_eC_e}(\lambda_1+\lambda_2)t^2+1+(\lambda_1+\lambda_2)t+ 
\frac{1}{2}(\lambda_1^2+\lambda_1\lambda_2+\lambda_2^2)t^2].
$$
Then, we conclude: for each $t$ in the interval $[0,\min \{-\frac{1}{\lambda_2};RC\} )$
 $$
 P_1(t)-Q_2(t)\le f(t)\le P_2(t)-Q_1(t).
 $$
 The quadratic polynomial 
 $$
P_1(t)-Q_2(t)=\frac{V}{R}(1-\frac{1}{RC}t)-
\frac{V}{R}[\frac{1}{R_eC_e}t+\frac{1}{2R_eC_e}(\lambda_1+\lambda_2)t^2+1+(\lambda_1+\lambda_2)t+ 
\frac{1}{2}(\lambda_1^2+\lambda_1\lambda_2+\lambda_2^2)t^2]
$$
has two nonnegative distinct roots: 
$$
t=0; \ t= 
\frac{2k}{C_e}\frac{1}{\frac{1}{R}(\frac{1}{C}+\frac{k}{C_e})^2+\frac{k}{R_eC_e^2}}
$$
and takes positive values on the interval $(0,
\frac{2k}{C_e}\frac{1}{\frac{1}{R}(\frac{1}{C}+\frac{k}{C_e})^2+\frac{k}{R_eC_e^2}}
)$ (because  
$P'_1(0)-Q'_2(0)=
k\frac{V}{R^2C_e}  >0$). Therefore $f(t)>0$ on the interval $(0,\tau)$, and the statement $(b)$ follows. $\Box$

\subsection{Remarks} 

\subsubsection{} 
\label{sec:explan}
The fact that 
$$
 f'(0)=
 k \frac{V}{R^2C_e} 
 $$ 
 in the proof of Theorem \ref{thelectrodgen} (generalizing $f'(0)=\frac{V}{R^2C_e}$ 
 in the proof of Theorem \ref{thelectrod}, when $k=1$), is significant. It can be interpreted as follows: a larger number of electrodes gives a stronger attenuating effect for the current. 
 
 \subsubsection{}  
The number $\tau$ in Theorem \ref{thelectrod} $(b)$, Theorem \ref{thelectrodgen} $(b)$ is obtained rigorously and it is explicit. It is an algebraic expression in the system parameters that is easy to calculate. This expression involves only 
elementary algebraic functions: rational functions and the square root function. We chose this approach having in mind 
a possibility of practical applications. 

To refine the discussion, one may ask about the smallest real number $\xi>0$ such that $f(\xi)=0$, either in section 
\ref{sec:mmodel} or in section \ref{sec:furan} (the exact value of $\xi$, not a lower bound for $\xi$). 
We are not going to pursue this direction in this paper, but here we will offer a brief comment. 
 It is a reasonable and mathematically interesting question. It translates to solving an equation of the form 
$\alpha_1 e^{\beta_1t} +\alpha_2 e^{\beta_2t} +\alpha_3 e^{\beta_3t}=0$ for real $t$ under appropriate assumptions.  
One can prove an existence statement using the Inverse Function Theorem. This gives  a suitably defined inverse function. In specific cases people refer to the Lambert function or its generalizations (e.g. the Lambert-Tsallis function). 
To give an easy example, the inverse function to $y(x)=xe^x$, $x\in[0,\infty)$, is the Lambert function (restricted to 
$[0,\infty)$). 
We acknowledge the work \cite{rms} where the values of Lambert-Tsallis function occur in the studies of the electrical circuits that lead to a linear combination of two exponential functions.

\section{Shielding}
\label{interf}

It has been suggested that shielding with Faraday fabric reduces or eliminates rheumatic pain \cite{farab1}. There is published clinical data on the effectiveness of this approach for muscle soreness  \cite{farab1},\cite{fb2}. 
Discussion in \cite{farab1} suggests 
possible reasons for this effect, specifically for joint pain. 

Consider a electric circuit that models the electrical activity in the joint, nerves, and surrounding tissue.
The inflamed tissue has higher conductance than normal tissue (\cite{karpuk} Fig. 7).
The inflamed tissue has higher capacitance too, compared to normal tissue (\cite{lee} Fig. 5). 
We leave it to the clinicians to decide what it means for a tender joint versus a healthy joint in terms of an expected response to the external electric fields.

Let us model shielding the joint with Faraday fabric by  
shielding the electric circuit with a hollow cylindrical conductor from an electromagnetic wave (a time-harmonic plane wave).  
To analyize this, we can use the results presented in 
section 4.3 of \cite{celozzi}. See also \cite{wutsai}. Consider the electric shielding coefficient 
$$
20\log\frac{|{\mathbf{E_0}}|}{|{\mathbf{E_S}}|}, 
$$
where ${\mathbf{E_0}}$ is the vector of the electric field on the axis of the cylinder in the absence of the shield and  ${\mathbf{E_S}}$
is the vector of the electric field on the axis of the cylinder in the presence of the shield. The magnetic shielding coefficient is defined similarly. 
From the conclusions in sec. 3 of \cite{wutsai}, sec. 4.3 of \cite{celozzi}, it follows that an appropriate choice of parameters results in  
significant attenuation of the fields.

\section{Circular conductor wire}
\label{sec:wifi}

In this section, we consider a mathematical model for one loop  or several loops of insulated copper wire (e.g. HDMI cable) placed on a person's head. Proposition  \ref{thwire} may be interpreted to consider this a way to relieve a certain type of   minor head pain.  

\begin{figure}
\begin{minipage}{0.32\textwidth}
\centering
\includegraphics[width=2.2in]{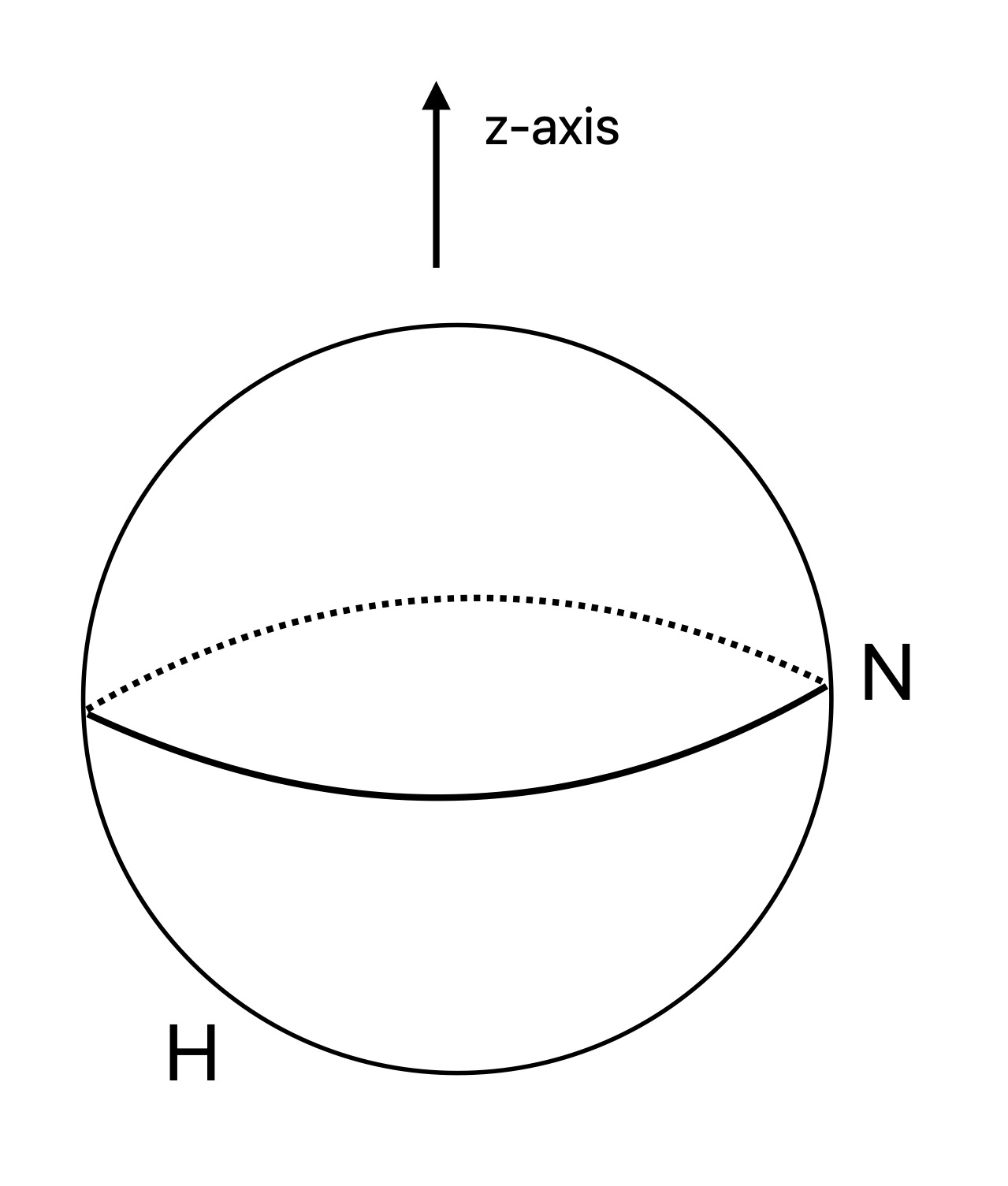}
\caption{A nerve is represented by the circle $N$ in the horizontal coordinate plane.} \label{figbraner}
\end{minipage}
\begin{minipage}{0.32\textwidth}
\centering
\includegraphics[width=2.2in]{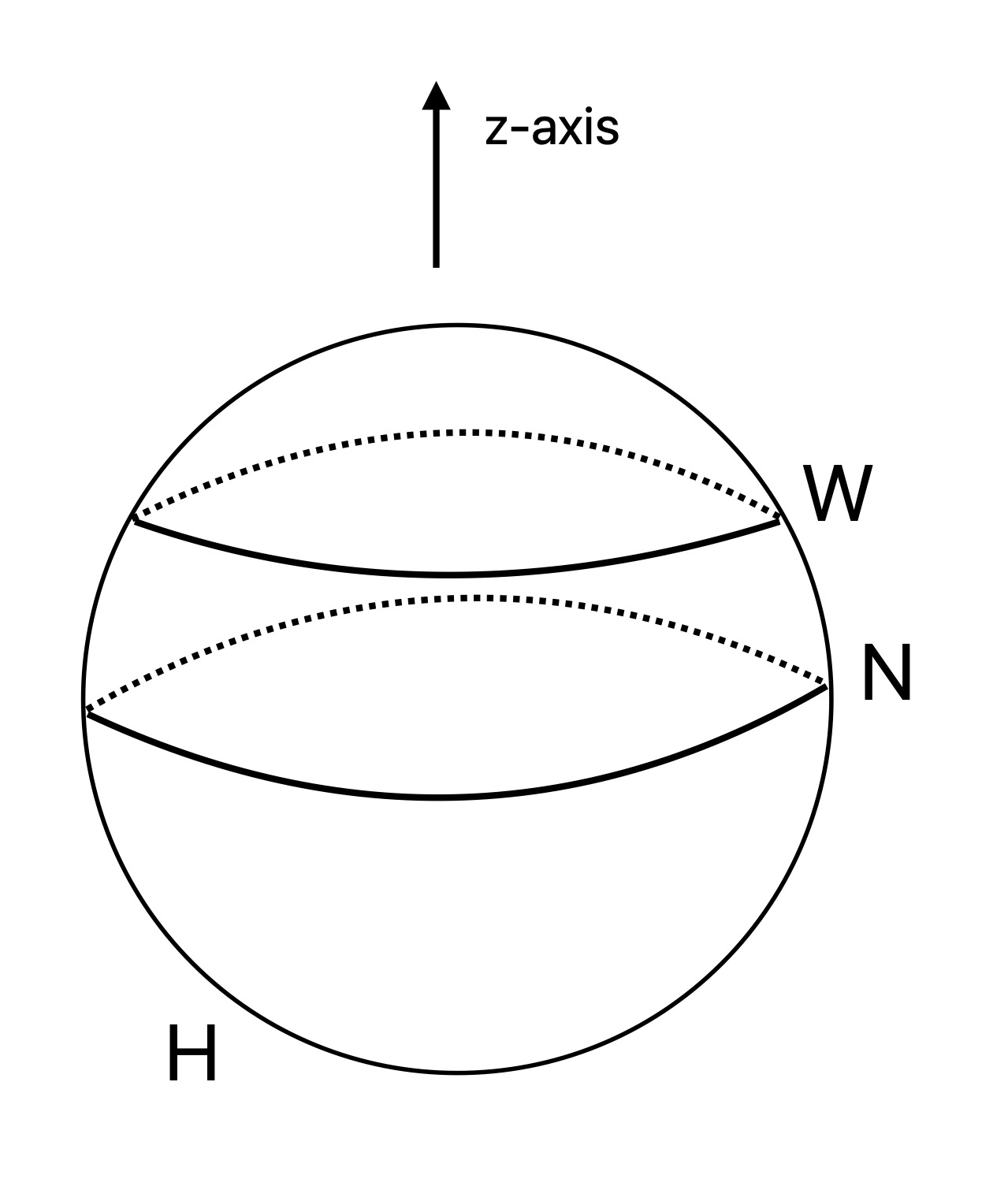}
\caption{The circular copper wire $W$ is placed above $N$.} \label{figbra}
\end{minipage}
\begin{minipage}{0.32\textwidth}
\centering
\includegraphics[width=2.2in]{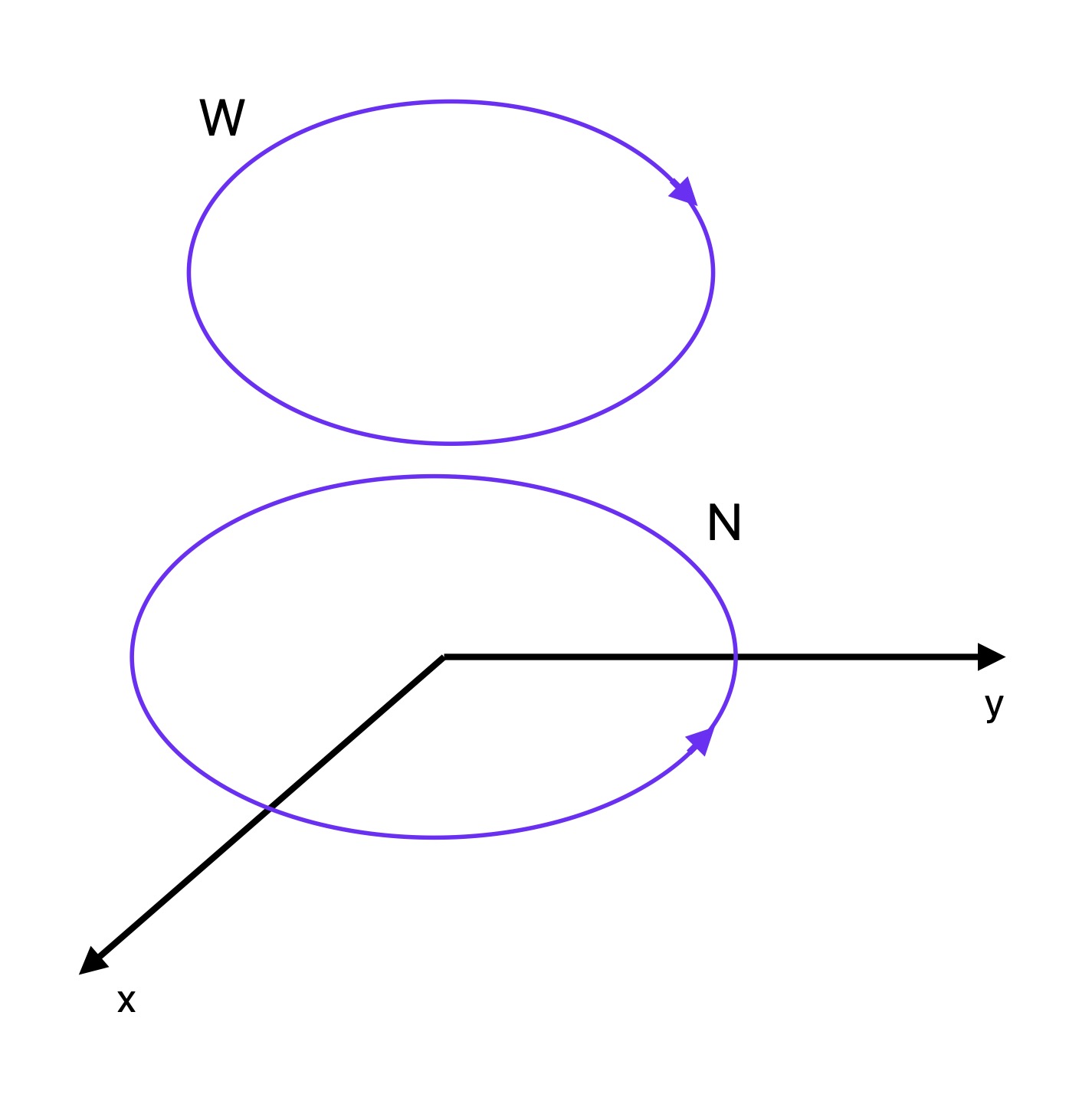}
\caption{Two loops: $W$, $N$.} \label{fig2lps}
\end{minipage}
\end{figure}

We will model the human head by a ball of radius $a>0$ in the $3$-dimensional space: 
$$
H=\{ (x,y,z)\in\R^3| \ x^2+y^2+z^2\le a^2\}. 
$$
Model a nerve by a conductor wire positioned on the big circle in the $xy$-plane
$$
N=\{ (x,y,z)\in\R^3| \ x^2+y^2+z^2= a^2; \ z=0\}
$$
(see Fig. \ref{figbraner}). 

Now, place a circular (insulated) copper conductor wire at a small distance above $N$:
$$
W=\{ (x,y,z)\in\R^3| \ x^2+y^2+z^2= a^2; \ z=\delta\}
$$
where $\delta>0$ is sufficiently small (see Fig. \ref{figbra}). 

If the electric current in the loop $N$, without loss of generality, is counterclockwise in the $xy$-plane (see Fig. \ref{fig2lps}), 
and increasing, then by the Lenz law the induced current in the loop $W$ is in the opposite direction.
The mutual inductance of two loops in this configuration is, by the formula 
(4.57) (sec. 4.4.1 \cite{paul})
$$
M=\frac{\mu}{2\sqrt{2}}a\sqrt{a^2-\delta^2}\int\limits_0^{2\pi}\frac{\cos\phi}{(a^2-a\sqrt{a^2-\delta^2}\cos\phi)^{1/2}}d\phi
$$
where $\mu$ is the permeability of medium.  Let $\varepsilon$ be the appropriate potential. We will allow it to be time-dependent: $\varepsilon=\varepsilon(t)$ a smooth function of $t$ on some interval $[0,T_{\varepsilon})$.  

Let $L_1$ be the self-inductance of $N$ and let $L_2$ be self-inductance of $W$. Let $R_1$, $R_2$ be the resistances.  
Before the cord is placed on the head, the electric circuit that represents $N$ only is in Fig. \ref{wirec1}.  
For this circuit, the differential equation and the initial condition are      
 $$
\left\{ 
\begin{array}{ll}  
\varepsilon-L_1\frac{dI}{dt}-IR_1=0 \\
I(0)=0
\end{array} \right.  \ .
$$
Without solving the initial value problem for $I(t)$, we immediately observe: 
\begin{equation}
\label{eq:deriv1}
I'(0)=\frac{\varepsilon(0)}{L_1}.
\end{equation}
We also observe that the solution of the initial value problem, for $t\ge 0$, is 
\begin{equation}
\label{itc1r}
I(t)=\frac{1}{L_1}e^{-\frac{R_1}{L_1}t}\int_0^t\varepsilon(u)e^{\frac{R_1}{L_1}u}du.
\end{equation}
In particular, if $\varepsilon$ is constant, then 
 $$
 I(t)=\frac{\varepsilon}{R_1}(1-e^{-\frac{R_1}{L_1}t}).
 $$

\begin{figure}[tb]
\centering
\begin{circuitikz}
\draw   (0,0)   coordinate(start)              
                to [short, i=$I$] ++ (1,0)    
                to [R=$R_1$ ]      ++ (3,0)    coordinate (a) 
                -- ++ (0,-1)
                to [L=$L_1$]    ++ (-3,0)
                -- ++ (-1,0)
                to [battery, l=$\varepsilon$, invert] (start)
        ;
\end{circuitikz}
\caption{One loop} \label{wirec1}
\end{figure}
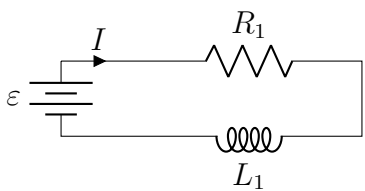

\begin{figure}
\begin{minipage}{0.32\textwidth}
\centering
\begin{circuitikz}
\draw   (0,0)   coordinate(start)              
                to [short, i=$I_1$] ++ (1,0)    
                to [R=$R_1$ ]      ++ (3,0)    coordinate (a) 
                -- ++ (0,-1)
                to [L=$L_1$]    ++ (-3,0)
                -- ++ (-1,0)
                to [battery, l=$\varepsilon$, invert] (start)
        ;
\end{circuitikz}
\caption{ } 
\label{figloop1}
\end{minipage}
\begin{minipage}{0.32\textwidth}
\centering
\begin{circuitikz}
\draw   (0,0)   coordinate(start)              
                to [short, i<=$I_2$] ++ (1,0)    
                to [R=$R_2$ ]      ++ (3,0)    coordinate (a) 
                -- ++ (0,-1)
                to [L=$L_2$]    ++ (-3,0)
                -- ++ (-1,0)
to (start)
        ;
\end{circuitikz}
\caption{} 
\label{figloop2}
\end{minipage}
\end{figure}

For two loops ($N$ and $W$, coupled, Fig. \ref{figloop1}/\ref{figloop2}), the system of equations is 
 $$
\left\{ 
\begin{array}{llll}  
\varepsilon=R_1I_1+L_1\dfrac{dI_1}{dt}-M\dfrac{dI_2}{dt} \\
0=R_2I_2+L_2\dfrac{dI_2}{dt}-M\dfrac{dI_1}{dt} \\
I_1(0)=0 \\
I_2(0)=0
\end{array} \right.  \ .
$$
Assume $L_1>0$, $L_2>0$,  $M<\sqrt{L_1 L_2}$, $R_1>0$, $R_2>0$.   
We conclude:
$$
I_1'(0)=\frac{\varepsilon(0) L_2}{L_1L_2-M^2}; \ I_2'(0)=\frac{\varepsilon(0) M}{L_1L_2-M^2},  
$$
and for $t\ge 0$ 
\begin{equation}
\label{eq:i2t}
I_2(t)=\frac{M}{(L_1L_2-M^2)(\lambda_1-\lambda_2)}\Bigl [ 
\lambda_1e^{\lambda_1 t}\int_0^t\varepsilon(u)e^{-\lambda_1u}du-
\lambda_2e^{\lambda_2 t}\int_0^t\varepsilon(u)e^{-\lambda_2u}du\Bigr ]
\end{equation}
(in particular, if $\varepsilon$ is constant, then 
$$
I_2(t)=
\frac{\varepsilon M}{(L_1L_2-M^2)(\lambda_1-\lambda_2)}(e^{\lambda_1t}-e^{\lambda_2 t}) \ )
$$
 where $\lambda_1$ and $\lambda_2$ are the two distinct negative solutions of the equation 
 $$
 (L_1L_2-M^2)\lambda^2+\lambda (R_2L_1+R_1L_2)+R_1R_2=0,
 $$ 
 that are 
 $$
 \lambda_1=\frac{-(R_2L_1+R_1L_2)+\sqrt{(R_2L_1+R_1L_2)^2-4R_1R_2(L_1L_2-M^2)}}{2(L_1L_2-M^2)}
 $$
 $$
 \lambda_2=\frac{-(R_2L_1+R_1L_2)-\sqrt{(R_2L_1+R_1L_2)^2-4R_1R_2(L_1L_2-M^2)}}{2(L_1L_2-M^2)}.
 $$ 
 The current $I_1$ is now expressed as  
 $$
 I_1(t)=\frac{\varepsilon}{R_1}-\frac{1}{MR_1}\Bigl ( L_1R_2I_2+(L_1L_2-M^2)\frac{dI_2}{dt}\Bigr ) ,
 $$
 with (\ref{eq:i2t}).
 
 Let $n$ be a positive integer. Let us extend our setup above, to replace $W$ (one loop) by $n$ closely placed insulated wire loops  
 (a closely spaced coil of $n$ turns: see the multiturn loop discussion in Sec. 4.2, 4.4.1 \cite{paul}). For $W$ with $n$ turns, denote the self-inductance, the mutual inductance, and the resistance by $\tilde{L}_2$, $\tilde{M}$,  $\tilde{R}_2$, respectively. Keep the notations ${L}_2$, ${M}$,  ${R}_2$, as above, for the one turn values. 
For practical purposes, $n$ will not be too large. If $n$ is small and the wire is thin and tightly wound, then we consider $N$ and $W$ as two coaxial coils in parallel horizontal planes, such that the distance between them in the vertical direction is small and the radia are close to each other. As $n$ gets larger, it is more reasonable to consider the second coil as a solenoid, and this gives us      
 $\tilde{L}_2=nL_2$, $\tilde{M}\approx nM$,  $\tilde{R}_2=nR_2$.   
We will now consider the following system of equations: 
  $$
\left\{ 
\begin{array}{llll}  
\varepsilon=R_1I_1+L_1\dfrac{dI_1}{dt}-nM\dfrac{dI_2}{dt} \\
0=nR_2I_2+nL_2\dfrac{dI_2}{dt}-nM\dfrac{dI_1}{dt} \\
I_1(0)=0 \\
I_2(0)=0
\end{array} \right.  \ 
$$ 
from which we immediately deduce: 
\begin{equation}
\label{eq:derivzer}
I_1'(0)=\frac{\varepsilon(0) L_2}{L_1L_2-nM^2}
\end{equation}
$$
 I_2'(0)=\frac{\varepsilon(0) M}{L_1L_2-nM^2}.  
 $$
As in the one turn case above, this system can be solved, to obtain $I_2(t)$ explicitly
 and then we find $I_1$ from 
 \begin{equation}
 \label{eq:i1tn}
 I_1(t)=\frac{\varepsilon}{R_1}-\frac{1}{MR_1}\Bigl ( L_1R_2I_2+(L_1L_2-nM^2)\frac{dI_2}{dt}\Bigr ) .
 \end{equation}
 \begin{proposition} 
 \label{thwire}
 Assume all notations are as above. 
 \noindent There is a positive integer $n_0$ such that for any integer $n\ge n_0$ there is 
 a real number $T_n>0$ such that $I(t)>I_1(t)$ for all $t\in (0,T_n)$, where 
 $I(t)$ and $I_1(t)$ are given by (\ref{itc1r}) and (\ref{eq:i1tn}) respectively. 
\end{proposition}
\noindent  {\bf Proof.} 
 The function 
 $$
 f(t)= I(t)-I_1(t)
; \ t\ge 0
 $$
 is $C^{\infty}$, \ $f(0)=0$. 
 From (\ref{eq:deriv1}) and (\ref{eq:derivzer})
 $$
 f'(0)=\frac{-\varepsilon(0)nM^2}{L_1L_2-nM^2} .
 $$
 Let $n_0\in\N$ be the smallest integer  such that  $n_0>\frac{L_1L_2}{M^2}$. Let $n$ be an integer such that 
 $n\ge n_0$. For this value of $n$,  $f'(0)>0$.  
 By continuity of $f'(t)$ it follows that $f'(t)>0$ in the interval $(0,T_n)$ for some $T_n$, meaning that $f$ is increasing in this interval, and since $f(0)=0$, it follows that $f(t)>0$ in $(0,T_n)$. This proves the statement. $\Box$
\begin{remark}
The fact that $I_1(t)<I(t)$ on some interval $(0,T_n)$ can be interpreted as follows:  
for small $t>0$, 
the configuration of the cable winded into $n$ turns decreases the current in the first loop (thus, weakens the signal). 
I.e. the current in $N$ when there is a multiturn $W$ nearby, positioned as described above, is weaker than the current in $N$ when $W$ is not present.  
\end{remark}

\section{Conclusions and discussion}

 In Section \ref{secextc}, we worked out a mathematical model  that suggests: placing two or more standard reference electrodes on the skin close to location of head pain would have a positive effect on a minor headache. Theorem \ref{thelectrod} (b) gives 
 a specific lower bound on the time over which this is guaranteed to occur, in terms of the parameters of the model. 
 Our conclusion that the effect is positive is based on the fact that the electric current in the nerve is reduced after the application 
 of the electrodes.     
 Theorem \ref{thelectrodgen} extends the conclusion of Theorem \ref{thelectrod} from two electrodes to several electrodes, and its proof shows that with a larger number of electrodes the effect is amplified (see the explanation in the section \ref{sec:explan}).

In Section  \ref{interf}, we stated that Faraday fabric eliminates pain in an aching joint, due to the shielding effect. This is consistent with published clinical evidence in medical literature (see the references in Section  \ref{interf}).   
 
 In Section  \ref{sec:wifi}, we examine up the mathematical model for the effect of several loops of a conductor wire on the electrical signal 
 in a nerve,  in the context of a minor headache. 
 The model is two coupled coaxial coils, positioned close to each other in parallel planes. In Proposition \ref{thwire},  we conclude that for small $t>0$,  adding  this second coil  (which has an appropriate number of turns), will decrease the current in the first loop (compared to the case when the first loop is by itself). This conclusion can be interpreted as a pain relieving effect.

 Throughout the paper,    we modeled the electrical signals in nerves, including pain signals, by electrical currents in cables.  Once we adopt this approach, 
we should take into consideration the effects of electromagnetic induction and interference.  To rephrase this less formally: 
propagation of electrical signals through the neuronal pathways that connect the brain with other parts of the human body is "communication via cable", and there is also a wireless component of the brain-body communication. The electrical currents in the brain neural network create an extracranial magnetic field (see the references in Section \ref{sec:intro} above for literature).  Thus, they act as an antenna transmitter. Loops in  the neural pathways outside the brain and in the close proximity to the brain act as an antenna receiver. In literature, similar  perspectives can be found, for example in \cite{kibret1}, \cite{kibret2}.

In our previous work \cite{barron}, \cite{barronkp}, we addressed the mathematical theory of signal propagation, having in mind applications to the networks of neurons. In this paper, we concentrate on mathematical models and practical approaches to applied problems.

\end{document}